\documentclass[letter, 10pt]{IEEEtran} 
\usepackage{graphicx}
\usepackage{float}
\usepackage{booktabs}
\usepackage{lipsum}
\usepackage{amsmath}
\usepackage{algorithmic}
\usepackage{mathtools}
\usepackage{stfloats}
\usepackage{caption}
\usepackage[linesnumbered,ruled,vlined]{algorithm2e}
\usepackage{epsfig}

\title{\LARGE \bf
Minimizing Electricity Cost through Smart Lighting Control for Indoor Plant Factories
}

\begin{document}

\author{\IEEEauthorblockN{Clement Lork\IEEEauthorrefmark{1}, Michael Cubillas\IEEEauthorrefmark{2}, Benny Kai Kiat Ng\IEEEauthorrefmark{1}, Chau Yuen\IEEEauthorrefmark{1}, Matthew Tan\IEEEauthorrefmark{3}}\\
	\IEEEauthorblockA{
		Engineering Product Development, Singapore University of Technology and Design, Singapore\IEEEauthorrefmark{1}\\
		3M APAC Pte Ltd, Singapore\IEEEauthorrefmark{2}\\
		James Cook University, Australia\IEEEauthorrefmark{3}\\
		Email: clement\_lork@sutd.edu.sg, cubillasm@asme.org, benny\_ng@sutd.edu.sg, yuenchau@sutd.edu.sg, matthew.tan@jcu.edu.au}
}

\maketitle


\begin{abstract}
Smart plant factories incorporate sensing technology, actuators and control algorithms to automate processes, reducing the cost of production while improving crop yield many times over that of traditional farms. This paper investigates the growth of lettuce (Lactuca Sativa) in a smart farming setup when exposed to red and blue light-emitting diode (LED) horticulture lighting. An image segmentation method based on K-means clustering is used to identify the size of the plant at each stage of growth, and the growth of the plant modelled in a feed forward network. Finally, an optimization algorithm based on the plant growth model is proposed to find the optimal lighting schedule for growing lettuce with respect to dynamic electricity pricing. Genetic algorithm was utilized to find solutions to the optimization problem. When compared to a baseline in a simulation setting, the schedules proposed by the genetic algorithm can achieved between 40-52\% savings in energy costs, and up to a 6\% increase in leaf area.
\end{abstract}

\section{INTRODUCTION}
Traditional agriculture across the world is challenged by population pressure, land scarcity, as well as abnormal weather conditions \cite{gebbers2010precision}. One of the more important agricultural advancements in the last 50 years is the concept of the plant factory. Plant factories can be built anywhere, and by stacking upwards, increases the agricultural output of a plot of land many times over traditional farms. Carefully controlled environments through the use of artificial lighting and air-conditioning allow crop production to thrive all year round, with reduced reliance on the outdoors climate when compared to traditional farming methods \cite{brandonnext}.

Smart sensors and machines are increasingly being integrated into plant factories with the development of Internet-of-Things technology. Through analysing the data collected from sensors, control algorithms can be implemented to automate processes and ultimately reduce the cost of production and to improve yield. Control algorithms for plant factories depend on the type of crops being grown and can range from simple rule-based control to the more complex optimization algorithms for decision making. \cite{chieochan2017iot} utilized humidity sensors and the NodeMCU microcontroller to track and control the humidity in a Lingzhi farm according to a rule-based flow chart. \cite{cruz2017design} developed a system to optimize the water and electrical resources of a aquaculture setup based on fuzzy logic control over a wireless sensor network with water level, pH and EC sensors. 

A major challenge of growing leafy vegetables in plant factories will be to manage the quantity and quality of artificial lighting to the crops \cite{kim2004green}, while balancing electricity cost. To optimize both plant growth and electricity cost, knowledge of how a plant develop with respect to different lighting or environmental conditions, as well as the power requirements of the artificial lighting, are required. Experiments will have to be conducted to collect such information if they are unavailable beforehand. An example is \cite{morimoto1996intelligent}, where the authors collected data on different tomato plants grown under different conditions, before compressing the information in a feed-forward neural network and optimizing for nutrient delivery schedule using a genetic algorithm.

Also, in recent years, countries like the US \cite{salari2016residential} and Japan \cite{murakami2018energy} are introducing dynamic electricity pricing to the electricity grid in an attempt to smooth out electricity demand, adding complexity to the electrical cost optimization problem for smart plant factories. 

As seen in the literature reviewed, deployment of sensing technology and control algorithms are conditional on the optimization objectives, crop grown, and farming setup. In this paper, we would like to explore the effects of red and blue LED lighting on lettuce (Latuca Sativa) grown in a Nutrient Film Technique (NFT) \cite{graves1983nutrient} hydroponics setup. Ultimately, we would also like to optimize the electrical cost of operating a lettuce farm subjected to the plant model and dynamic electrical pricing. Our methodology is summarized in three steps: raw plant growth data collection, plant growth modelling, and finally lighting cost optimization.

The paper made the following contributions:
\begin{itemize}
    \item For plant growth data collection, we proposed an experimental IOT setup and image processing framework to monitor the leaf area growth of lettuce
    \item For plant growth modelling, we modelled the effects of LED lightning on lettuce in a neural network
    \item For lighting cost optimization, we optimized the lighting schedule for operation based on plant model and dynamic electricity pricing
\end{itemize}

While the this paper focuses on optimizing the lightning control of indoor plant factories, the same methodology can be applied to other operational factors like temperature, humidity, and nutrients control. The rest of the paper is organized as such: Section II discusses the experimental setup and data processing methodology, Section III introduces the plant growth model, Section IV describes the lighting cost optimization algorithm and presents the outcome of the optimization. Finally Section V is the conclusion and discusses future work.

\section{SETUP AND IMAGE PROCESSING FRAMEWORK}
\subsection{Experimental Setup}
\begin{figure}[htb!]\centering \footnotesize
	\includegraphics[width=80mm]{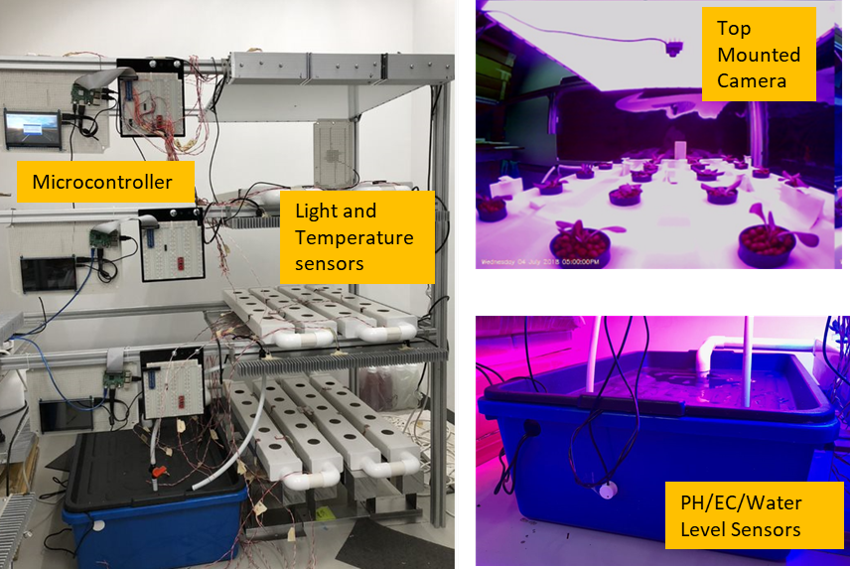}
	\caption{Experimental setup of NFT hydroponics system}
	\label{setup}
\end{figure}

\begin{figure}[htb!]\centering \footnotesize
	\includegraphics[width=80mm]{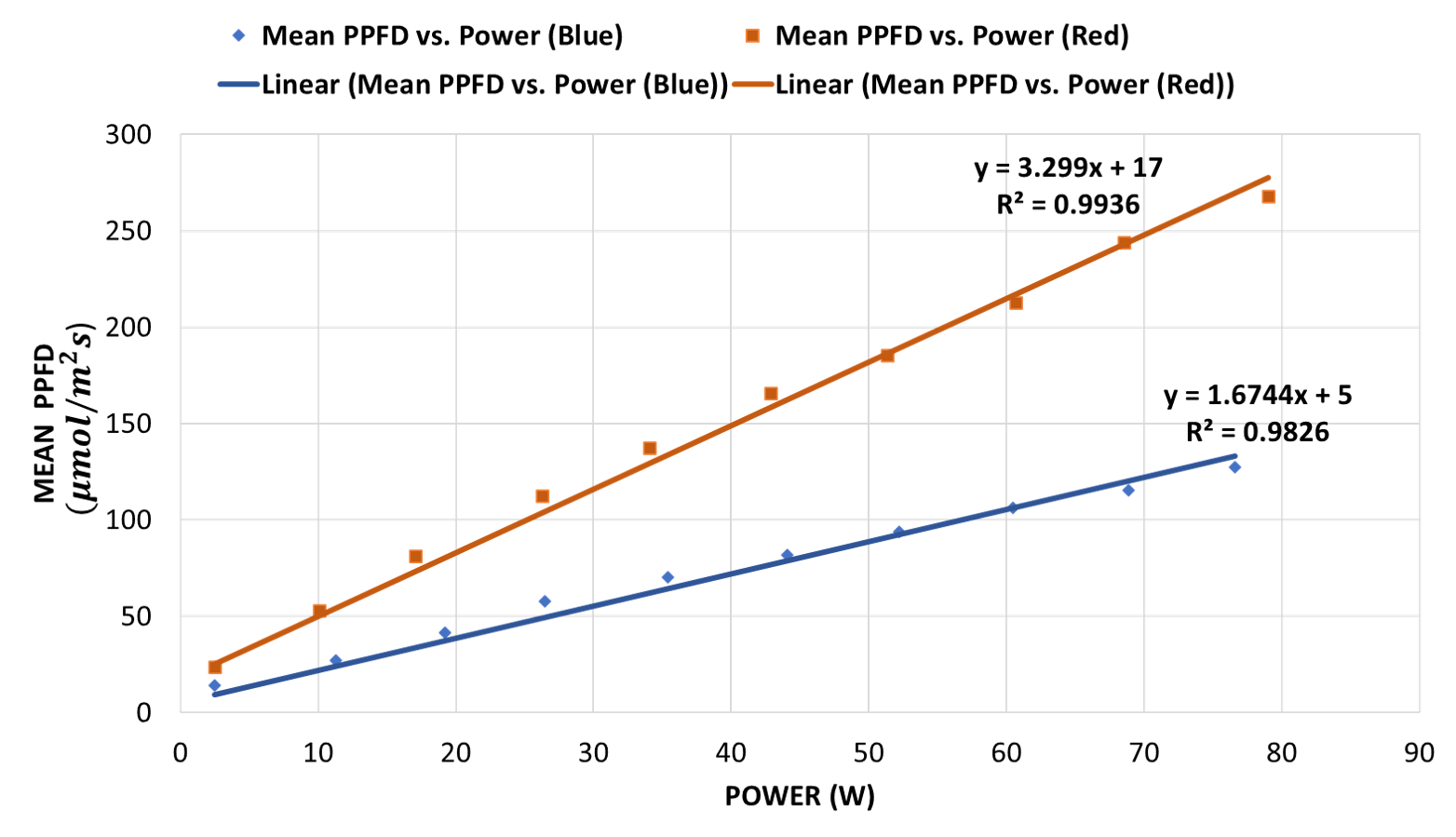}
	\caption{Power requirements of LED lighting panel}
	\label{power_req}
\end{figure}

The experimental setup shown in Fig. \ref{setup} is housed in an air conditioned room where the temperature is kept at 25 $^{\circ}$deg Celsius. It consists of three levels of NFT hydroponics grow-bed, with the lighting conditions of each level individually controlled with a Raspberry Pi 3 micro-controller. The light source for each level is a dimmable panel with integrated red (650-670nm) and blue (450-465nm) LEDs mounted at 40cm above the grow-bed. Each grow-bed can support up to 20 lettuce plants, sharing the same nutrient solution with the other levels. The nutrient solution is circulated through the grow-beds with a pump from a 120 litre tank. The aim of the experiment is to collect information on how the lettuce plants will react to changes in lighting conditions. Hence, the values for nutrient concentration as well as the pH, as measured by the electro-conductivity (EC) sensor and pH sensor sourced from Atlas Scientific, are kept at consistent between 1600-2000 $\mu S/cm$ and $6.4-6.7$. Once the EC, pH, or water-level values dip below the desired threshold, a notification to top up the nutrient solution will be sent out by the micro-controller.
Lighting, temperature and humidity conditions on each test bed are monitored by an array of sensors placed on each test bed. The raw values for the ISL29125 lux sensors used to monitor lighting were calibrated to the standard $\mu mol/m^2s$ units used to measure photosynthetic flux density (PPFD). Temperature and humidity under each were measured by DHT22 sensors. Data from the experimental setup are uploaded to a web server hourly. Prior to seeding of the lettuce plants, the power requirement of the LED panel at each red and blue PPFD settings were quantified in Watts using a power meter. The power requirements are shown in Fig. \ref{power_req}.

\subsection{Image Processing}
A top mounted camera is utilized to monitor the plant area throughout the course of each experiment in an attempt to quantify plant growth, as there is a strong correlation between plant leaf area and plant biomass  \cite{weraduwage2015relationship}. Pictures on the growth of the plant were also sent to a web server hourly and the images batch-processed at the end of the experiment. However, it is a challenge for computers to recognize the leaf area for plants in an image. Leaf area segmentation techniques include simple thresholding like Otsu thresholding \cite{kurita1992maximum}, to state of the art convolutional neural network (CNN) based approaches \cite{atanbori2018towards}. In our use case, simple thresholding does not work well with our images as they have too much variations in color, while CNN based techniques require too much effort to retrain the network for our setup. To avoid the time consuming task of manually labelling each image, we modified the K-means algorithm in \cite{dhanachandra2015image} to come up with a helper algorithm in Algorithm \ref{segmentation} for plant segmentation. 

\begin{algorithm}
Undistort raw images via homography estimation \cite{dubrofsky2009homography}\\
Cluster images using K-means algorithm via their pixel brightness and distance from center of each pot\\
Estimate the threshold to determine which clusters are plant clusters and which are non-plant clusters \linebreak
(I) Sample a few images in chronological order from a experimental sequence \linebreak
(II) Manually determine which clusters are part of the plant and find the centroid of the cluster which can still be classified as a plant cluster \linebreak
(III) (Assuming plant grow exponentially) Fit an exponential rule to predict the growth of the maximum centroid from data gather in (I) and (II) \\
Apply the exponential rule to all images in the growth stage to obtain segmented images. Final results are shown in Fig. \ref{seg_result}
\caption{Batch Image Segmentation of Leaf Area}
\label{segmentation}
\end{algorithm}

\begin{figure}[htb!]\centering \footnotesize
    \includegraphics[width=85mm]{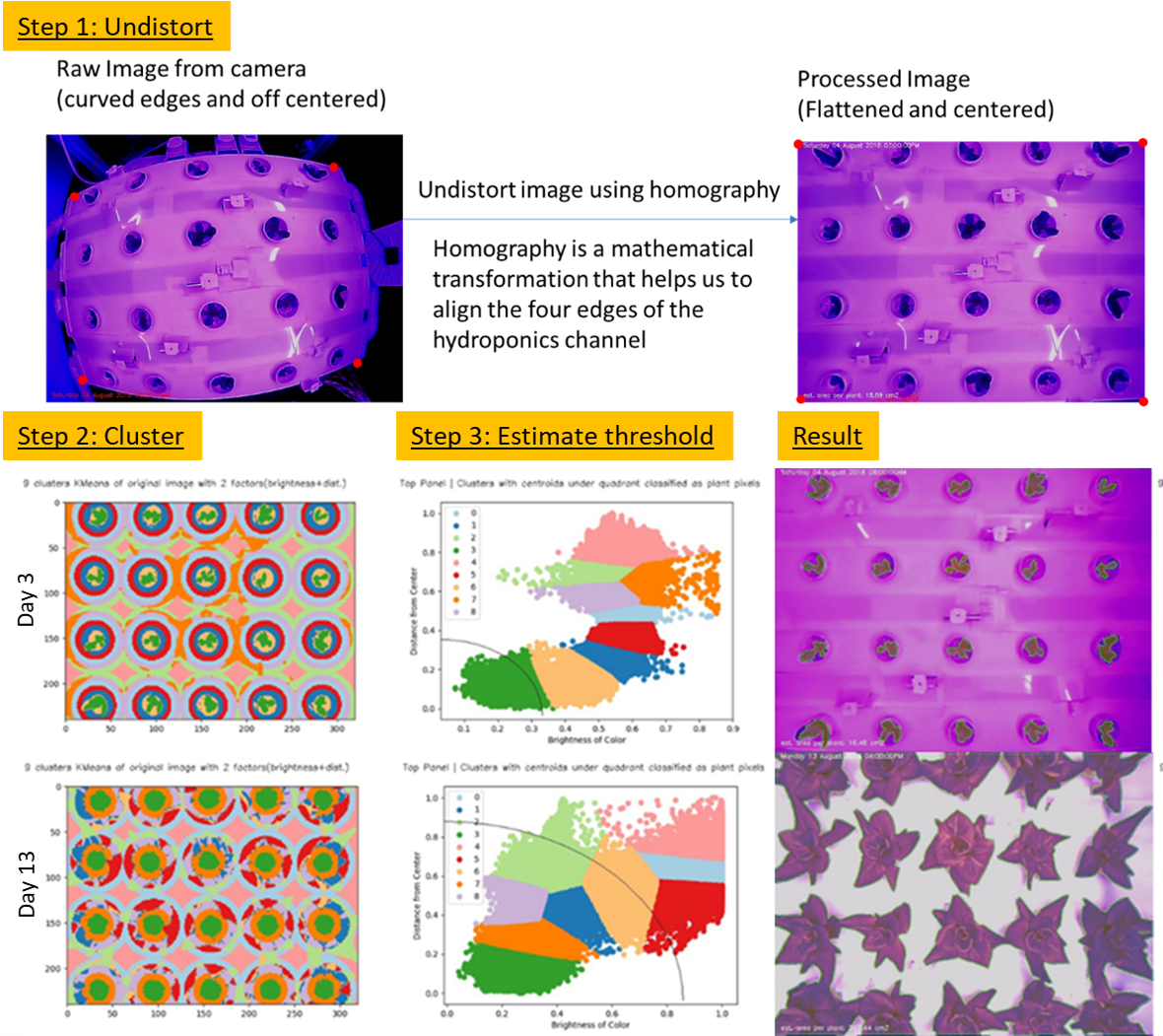}
	\caption{Results of Algorithm \ref{segmentation}}
	\label{seg_result}
\end{figure}

In Fig. \ref{seg_result}, under step 2 to step 3, we show the results of applying Algorithm \ref{segmentation} to images taken from day 3 to day 13 after the lettuce plants are transplants into the setup.

\section{PLANT GROWTH MODELLING}%

\begin{table}[]
\caption{Settings for different experiments}
\label{exp_set}
\resizebox{\columnwidth}{!}{
\begin{tabular}{@{}llll@{}}
\toprule
Experiment &
  \begin{tabular}[c]{@{}l@{}}Avg. Red PPFD On\\ $(\mu mol/m^2s)$ \\\end{tabular} &
  \begin{tabular}[c]{@{}l@{}}Avg. Blue PPFD On\\ $(\mu mol/m^2s)$\end{tabular} &
  \begin{tabular}[c]{@{}l@{}}Duty Cycle \\ from 0000hrs\end{tabular} \\ \midrule
1 & 166 & 18 & 18hrs On 6hrs Off  \\
2 & 144 & 54 & 18hrs On 6hrs Off  \\
3 & 160 & 54 & 9hrs On 3hrs Off   \\
4 & 87 & 37 & 13hrs On 11hrs Off \\ \bottomrule
\end{tabular}}
\end{table}

\subsection{Plant Growth Equation}
The relative growth rate (RGR) of a plant's biomass typically follows an exponential relationship \cite{klassen2004real}, as in Eq. \ref{eq1}.

\begin{equation} \label{eq1}
RGR  = ln (M_2 / M_1)/\triangle t
\end{equation}
where $M_2$ is the final dry mass, $M_1$ is the initial dry mass and $\triangle t$ is the change in time in days.
In our experiment, we assumed that the \textbf{increase in leaf area} of a plant is related exponentially to a non-linear function that is dependent on the averaged light energy given to the plant over $\triangle t$, and external variables like EC and pH. The initial leaf area $L_1$ and final leaf area $L_2$ takes the place of $M_1$ and $M_2$, resulting in Eq. \ref{eq2}. $t$ is the time in days since transplanting.

\begin{equation} \label{eq2}
\begin{split}
& f(PPFD_{red_{avg}},PPFD_{blue_{avg}},EC,pH,t) \\
& = ln (L_2 - L_1)/\triangle t
\end{split}
\end{equation}

\subsection{Plant Model}
\begin{figure}[htb!]\centering \footnotesize
    \includegraphics[width=80mm]{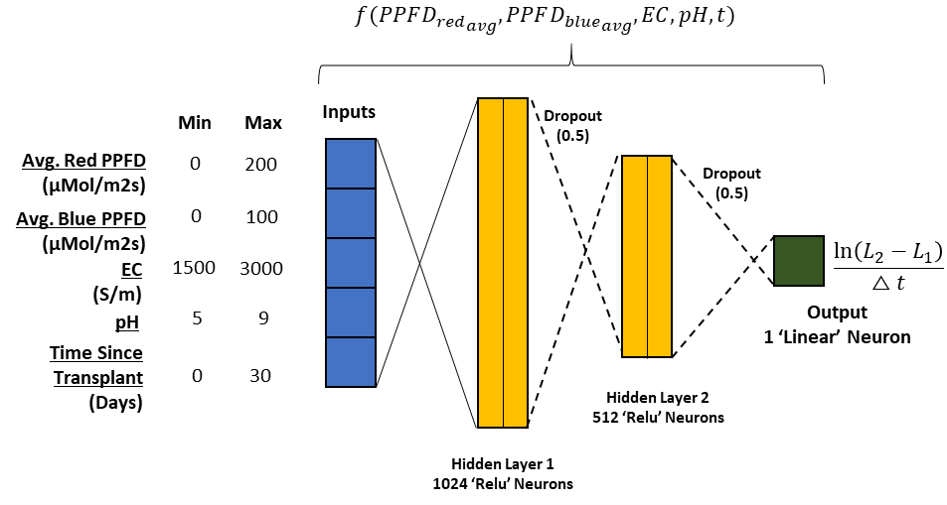}
	\caption{Structure of Neural Network}
	\label{nn}
\end{figure}

A neural network is a universal function approximator \cite{sifaoui2008use}, and will allow us to approximate the relationship between variables in Eq. \ref{eq2}. Data from the 4 experiments are split into testing and training data. Data from experiments 1,2,4 goes into the training set, while data from experiment 3 are used for testing. The design of the neural network, as well as the variables used are shown in Fig. \ref{nn}. Variables are min-maxed to 0-1 according to the range displayed, again in Fig. \ref{nn}. The neural network is a two layer feed forward network with 'relu' neurons. Dropout of probability 0.5 is used for regularization to improve the accuracy of the network. The network was trained with back-propagation using the Adam optimizer with Mean Squared Error as the loss function for 2000 epochs.

A simple linear regression is also used to approximate $f(PPFD_{red_{avg}},PPFD_{blue_{avg}},EC,pH,t)$ for comparison with the neural network. The testing and training errors for both methods are summarized in Table \ref{error_tab} in terms of Mean Squared Error (MSE) and $R^2$ values to determine the goodness of fit of the model. The model is trained on the training data, and tested with either the training data or the testing data.

\begin{table}[htb!]
\caption{Training and testing error for plant model}
\label{error_tab}
\begin{tabular}{@{}lll@{}}
\toprule
\textbf{\begin{tabular}[c]{@{}l@{}}Predicted Leaf Area vs.\\ Actual Leaf Area\end{tabular}} &
  \textbf{\begin{tabular}[c]{@{}l@{}}Linear Regression\\ (MSE)\end{tabular}} &
  \textbf{\begin{tabular}[c]{@{}l@{}}Neural Network\\ (MSE)\end{tabular}} \\ \midrule
\textbf{Training} & 1513.919 & \textbf{126.3719} \\
\textbf{Testing}  & 2017.402 & \textbf{1807.548} \\ \midrule
\textbf{\begin{tabular}[c]{@{}l@{}}Predicted Leaf Area vs.\\ Actual Leaf Area\end{tabular}} &
  \textbf{\begin{tabular}[c]{@{}l@{}}Linear Regression\\ ($R^2$)\end{tabular}} &
  \textbf{\begin{tabular}[c]{@{}l@{}}Neural Network\\ ($R^2$)\end{tabular}} \\ \midrule
\textbf{Training} & 0.8601   & \textbf{0.9902}   \\
\textbf{Testing}  & 0.8783   & \textbf{0.8864}   \\ \bottomrule
\end{tabular}
\end{table}

The neural network model performed much better than that of the linear regression by having a lower MSE and high $R^2$ value on both the training and testing datasets. However the difference in accuracy is lesser in the testing dataset as compared to the training dataset due to a slight distribution shift in the testing dataset. 

\begin{figure}[htb!]\centering \footnotesize
    \includegraphics[width=72mm]{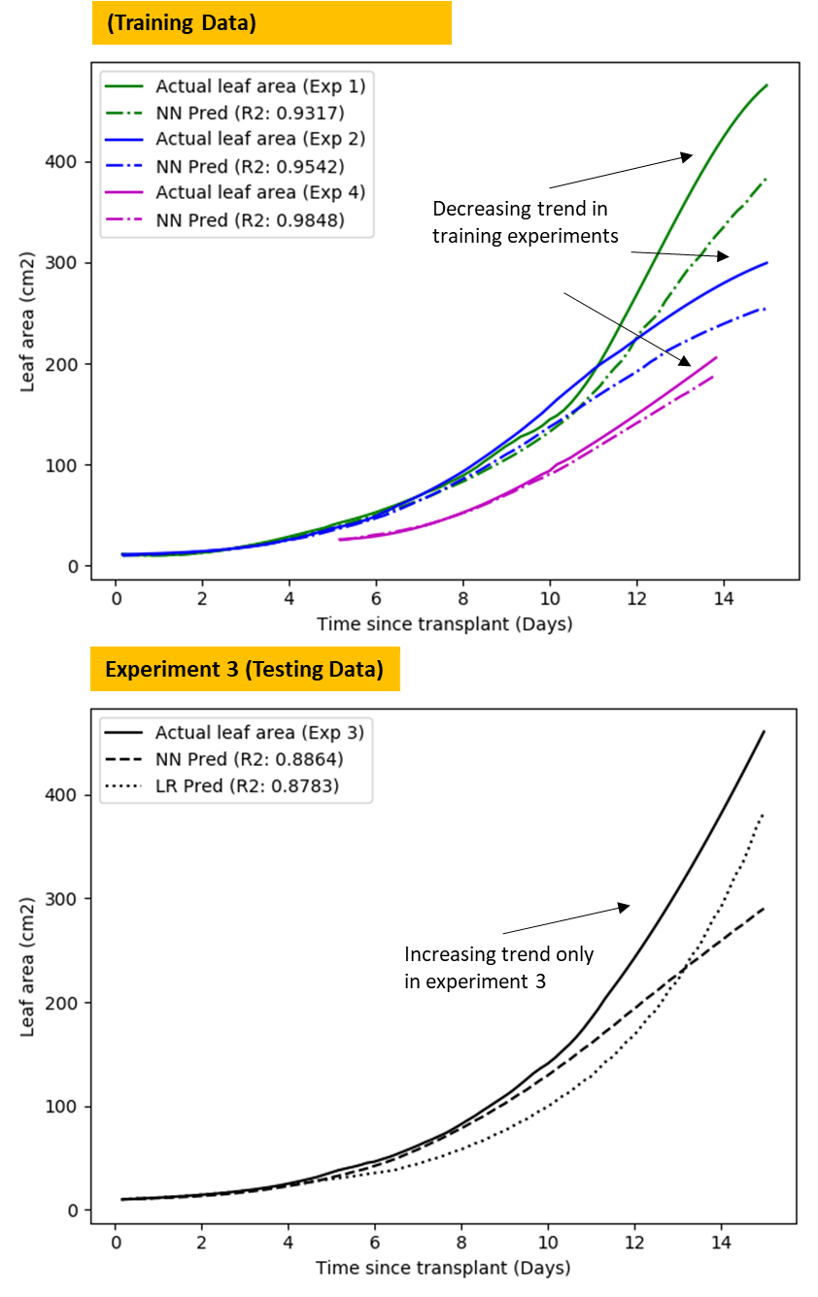}
	\caption{Leaf area prediction}
	\label{pred_result}
\end{figure}

Fig. \ref{pred_result} shows the shift in distribution and results of the eventual leaf area prediction. With more data, the neural network is expected to perform better than that of the linear regression, with its ability to model the non linear behaviour that may arise with lighting and plant growth interaction .

\subsection{Sensitivity Analysis}
\begin{figure}[htb!]\centering \footnotesize
    \includegraphics[width=75mm]{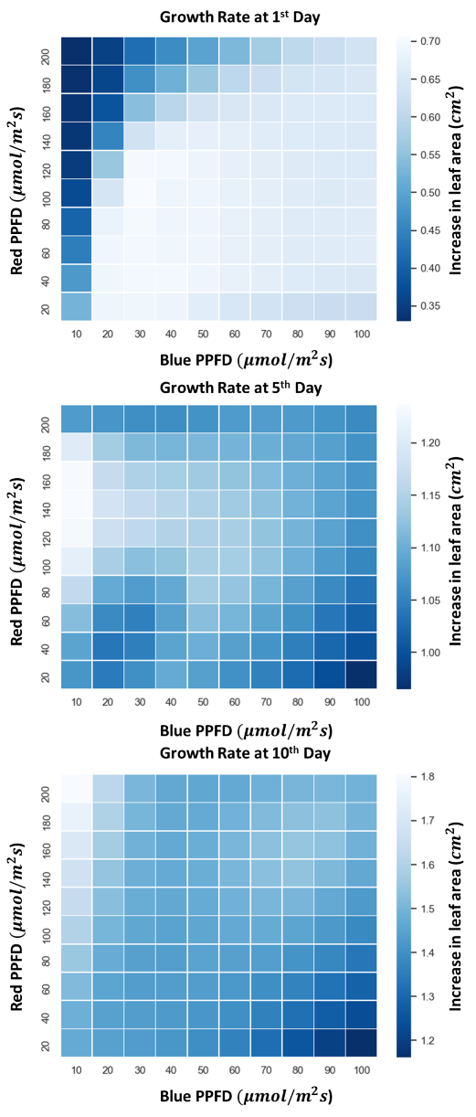}
	\caption{Growth rate of lettuce with different lighting settings at different stages}
	\label{growth}
\end{figure}

Once we have the plant model, we can track the sensitivity of the increase in leaf area to changes in lighting conditions. Eq. \ref{eq2} can be manipulated as such to Eq. \ref{eq3}:

\begin{equation} \label{eq3}
\begin{split}
L_2-L_1 = e^{f(PPFD_{red_{avg}},PPFD_{blue_{avg}},EC,pH,t) \cdot \triangle t}
\end{split}
\end{equation}
Assuming $EC=1800$, $pH=6.5$, and $\triangle t=1 hr$, we investigated the increase in leaf area ($L_2-L_1$) when $t$= $1^{st} day$, $5^{th} day$, and $10^{th} day$ across the range of the light settings. The results are shown in Fig. \ref{growth}.

The brighter portions of the heatmap in Fig. \ref{growth} denotes the lighting settings with higher growth rates. As we see, at the early stages of lettuce growth, the plant prefers a mixture of red and blue lighting for best growth. After some time, the preference of the plant switches over to red light. This is consistent in the findings in \cite{zhang2018effects}, who recommends mixing red and blue light in the ratio of 2.2:1 for lettuce growth, in the case of non adjustable LEDs. This ratio might not be the most cost effective option as blue leds will require more energy to provide the same irradiance level as red leds (Fig. \ref{power_req}). For LEDs with adjustable levels, we could leverage on the natural preference of the plant and adjust the lighting conditions for maximum growth and savings at each stage. For best growth on the 5th and 10th day, the ideal red to blue ratio is 14:1 and 20:1 respectively, as suggested by the plant model.

\section{LIGHTING COST OPTIMIZATION}
 One of the concerns for indoor plant factories is electricity pricing, due to the reliance on artificial lighting to drive plant growth. In countries where there are dynamic electricity prices, this can be a bane or a boon. For the growth of vegetative plants which usually do well under a 24 hr lighting schedule, the lighting cycle will have to be planned away from periods when electricity cost are high to prevent huge running cost. At the same time if plant factories can take advantage of the periods of lower-ed electricity costs, they could experience savings in electricity bills. The problem will be even more interesting if solar photovoltics are involved, although this aspect is not investigated in this paper. Prerequisites to optimizing the cost of electricity in a plant factories while maximizing the biomass of the plant depends on three factors: knowledge of the electricity prices, knowledge of how plants react to lighting, and knowledge of the energy consumption of the lighting system. 
 In our case study, we attempt to optimize the energy consumption and maximize the biomass of lettuces plants in our setup for a period of 15 days. 
 The electricity price plan for reference is taken from the dynamic pricing scheme of Tokyo Electric Power Company \cite{ida2015electricity} and listed in Table \ref{pricing}. The plant model is described in Eq. \ref{eq2} in Section. III and the operating characteristic of the LED lighting panel is illustrated in Fig. \ref{power_req}. This results in the following optimization problem to maximise the profit margin of operation for a single lighting panel and grow-bed:
\begin{equation}
\label{eq4}
\resizebox{80mm}{!}{$
\begin{aligned}
&\underset{s}{max}& P.L_{360}.n - \sum_{\delta=0}^{\delta=360} E(\delta,P(s_\delta)) \\
&\text{s.t.} & L_{\delta+1} - L_{\delta}& = \text{Plant Model in Eq. \ref{eq3}}  \\
& & E(\delta,P(s_\delta)) & = \text{Elec. cost in Table \ref{pricing}} \\
& & P(s_\delta) & =  \text{LED power in Fig. \ref{power_req}} \\
& & L_{360} & > 400
\end{aligned}$}
\end{equation}

In this problem, the time step $\delta$ represents 1 hr in a day, resulting in 360 total timesteps over a period of 15 days.The control sequence $s$ is a 2 by 360 vector, representing the red and blue lighting level at each hour for a period of 15 days.  $L_t$ is the average leaf area of a lettuce at time $\delta$ and $n$ is the number of lettuce cultivated per panel, which is 20. $P$ is the cost of lettuce per unit area per lettuce and will determine the output of the optimization. If P is very low, the optimal behavior of the system will be to not turn on any lights, which is detrimental to the growth of the plants. To prevent this, we also set a constraint on the size of the lettuce in the optimization problem to ensure it is at least 400 $cm^2$ at the end of the 15 days.

\begin{table}[h]
\caption{Dynamic Electricity Pricing in Japan}
\label{pricing}
\begin{tabular}{@{}l|lllll@{}}
\toprule
Time of Day (hrs) &
  \begin{tabular}[c]{@{}l@{}}0000-\\ 0700\end{tabular} &
  \begin{tabular}[c]{@{}l@{}}0700-\\ 1000\end{tabular} &
  \begin{tabular}[c]{@{}l@{}}1000-\\ 1700\end{tabular} &
  \begin{tabular}[c]{@{}l@{}}1700-\\ 2300\end{tabular} &
  \begin{tabular}[c]{@{}l@{}}2300-\\ 0000\end{tabular} \\ \midrule
Cents (Yen) / kWh &
  12 &
  25 &
  38 &
  25 &
  12 \\ \bottomrule
\end{tabular}
\end{table}

\subsection{Genetic Algorithm}
The optimization problem in Eq. \ref{eq4} is a non-linear one due to the presence of non-linear constraints. Among the many other ways to solve non-linear optimization problems, Genetic Algorithm (GA) is reputed as a fast and efficient solver if the fitness function can be quickly evaluated, making it a great algorithm to solve Eq. \ref{eq4}. The first step in a genetic algorithm is to represent the solution of the problem in a chromosome. For our optimization problem, the chromosome is a 360 long vector with each bin taken up by a tuple, $(r_{\delta},b_{\delta})$. Both $(r_{\delta},b_{\delta}) \in Z:\{1 ... 10\}$, with 1 representing 10\% of maximum PPFD and 10 representing 100\% of max PPFD. The maximum PPFD for red LEDs in our system is 200 $\mu mol/m^2s$, while for the blue LEDs it is at 100 $\mu mol/m^2s$. A $(r_{\delta}=3,b_{\delta}=3)$ tuple will mean that the red LEDs are outputting PPFD at 60 $\mu mol/m^2s$ while the blue LEDs are outputting at 30 $\mu mol/m^2s$. The fitness function is the objective function in Eq. \ref{eq4}. Beyond that, the steps in GA like population generation, crossover, and mutation follows that of the canonical version stated in \cite{whitley1994genetic}. Parameters relevant to GA are listed in Table \ref{ga}.

\begin{table}[h]
\caption{GA parameters}
\label{ga}
\begin{tabular}{@{}lllll@{}}
\toprule
\begin{tabular}[c]{@{}l@{}}Population\\ Size\end{tabular} &
  Parents &
  \begin{tabular}[c]{@{}l@{}}Crossover\\ Points\end{tabular} &
  \begin{tabular}[c]{@{}l@{}}Mutations \\ per child\end{tabular} &
  Generations \\ \midrule
100 &
  50 &
  6 &
  20 &
  300 \\ \bottomrule
\end{tabular}
\end{table}

\subsection{Results and Discussions}
Different values of $P$ were tested, $P$ = 0, $P$ = 0.001, and $P$ = 0.01. The generated control sequences from GA were also compared to a $(r_{\delta} = 7,b_{\delta} =7)$ 24hr always-on baseline sequence, as recommended by \cite{zhang2018effects}. 

\begin{figure}[htb!]\centering \footnotesize
    \includegraphics[width=75mm]{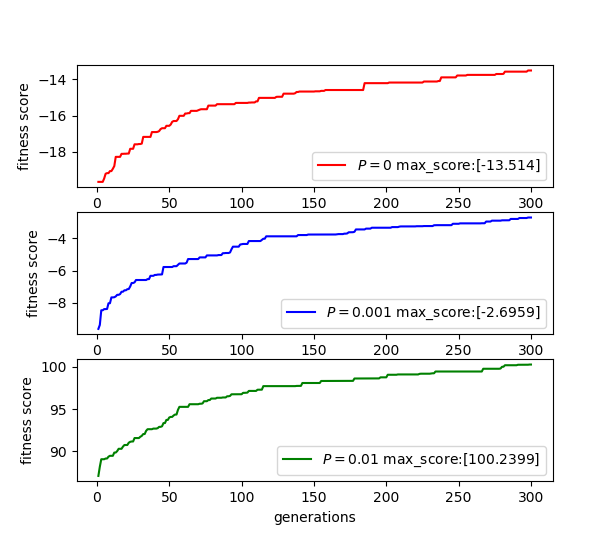}
	\caption{Convergence characteristics of GA for different $P$ values}
	\label{convergence}
\end{figure}

Fig. \ref{convergence} shows the convergence of the fitness function in various cases of $P$. 

When $P$ = 0, the fitness function is dominated by the cost of electricity over the period of 15 days. Hence, this will lead to the control sequences being optimized for a low electricity cost, subjected to constraint that the final leaf area of the plant must be more than 400 $cm^2$. When $P$ is low at 0.001, the revenue from selling the plant will be approximately equal to the cost of electricity, this will induce the GA to search for control sequences that balances the cost of electricity with the final leaf area. Finally when $P$ is high at 0.01, the GA will tend to search for control sequence that favor a larger final leaf area over lowering the cost of electricity.

\begin{figure}[htb!]\centering \footnotesize
    \includegraphics[width=75mm]{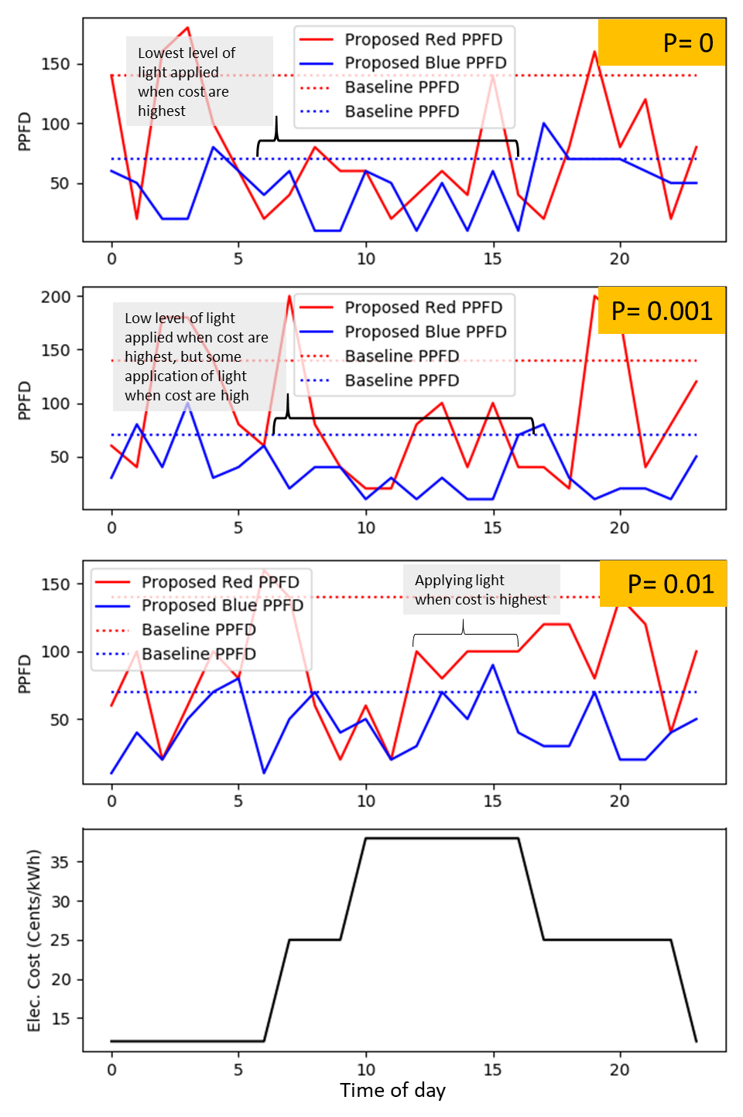}
	\caption{Control sequences for different $P$ values (first 24 hours)}
	\label{control_seq}
\end{figure} 

The control sequence for the $1^{st}$ 24 hour sequence in $s$ for various $P$ values are shown in Fig. \ref{control_seq}. We see that when the $P$ is highest, the GA optimizes the growth of the plant by offering more light even when the cost of electricity is high. When $P$ is the lowest, the GA choose to only operate the lights at the portion when electricity prices are the lowest. For $P$ = 0.001, the GA balances the growth of the plant with the cost of electricity.

The final leaf area and electricity cost are summarized in the Table. \ref{comparision}. Profit is the objective function of the problem defined in Eq. \ref{eq4}. When compared to the baseline, the GA allows the system to achieve between 40-52\% savings in electricity cost. Meanwhile, the leaf area output can be improve by up to 6\%. 
\begin{table*}[]
\caption{Leaf area and electricity cost comparison for GA with different $P$ values}
\label{comparision}
\resizebox{160mm}{!}{
\begin{tabular}{@{}lllllllllll@{}}
\toprule
\begin{tabular}[c]{@{}l@{}}Plant Price\\ (cents/$cm^2$\\ /plant)\end{tabular} &
  \begin{tabular}[c]{@{}l@{}}Baseline\\ Leaf Area\\ ($cm^2$)\end{tabular} &
  \begin{tabular}[c]{@{}l@{}}Baseline\\ Elec. Cost\\ (cents)\end{tabular} &
  \begin{tabular}[c]{@{}l@{}}Proposed\\ Leaf Area\\ ($cm^2$)\end{tabular} &
  \begin{tabular}[c]{@{}l@{}}Proposed \\ Elec. Cost\\ (cents)\end{tabular} &
  \begin{tabular}[c]{@{}l@{}}Proposed \\ revenue\\ from plants\\ (cents)\end{tabular} &
  \begin{tabular}[c]{@{}l@{}}Profit-Proposed\\ (cents)\end{tabular} &
  \begin{tabular}[c]{@{}l@{}}Profit-Baseline\\ (cents)\end{tabular} &
  \textbf{\begin{tabular}[c]{@{}l@{}}\%\\ Improvement\\ in Leaf Area\end{tabular}} &
  \textbf{\begin{tabular}[c]{@{}l@{}}\%\\ Improvement\\ in Elec. Cost\end{tabular}} &
  \textbf{\begin{tabular}[c]{@{}l@{}}\%\\ Improvement\\ in Profits\end{tabular}} \\ \midrule
\textit{P=0} & 458.0541 & 27.9206 & 442.3375 & 13.5139 & 0        & -13.5139 & -27.9206 & \textbf{-3.4311} & \textbf{51.5985} & -                \\
P=0.001      & 458.0541 & 27.9206 & 452.5732 & 13.5576 & 10.8617  & -2.6958  & -16.9273 & \textbf{1.1965}  & \textbf{51.4422} & -                \\
P=0.01       & 458.0541 & 27.9206 & 486.6086 & 16.5461 & 116.7860 & 100.2399  & 82.0128  & \textbf{6.2338}  & \textbf{40.7387} & \textbf{22.2247} \\ \bottomrule
\end{tabular}}
\end{table*}

\section{CONCLUSION AND FUTURE WORK}
In all, we have observed the growth of lettuce (Latuca Sativa) in our IOT-based experimental setup, and collected the data for analysis. A image processing framework based on the K-means method is used to extract the leaf area information from images captured during the experimental phase. Using the leaf area information and sensor data from the experiments, we modelled the growth of the plant with respect to changes in lighting conditions, pH and EC factors. Finally, we proposed an optimization problem to find the best lighting schedule to reduce cost of production based on our plant growth model. Genetic Algorithm is used to find solutions to the optimization problem. The optimized lighting system is able to achieve between 40-52\% savings in electricity cost while improving the leaf-area of the plants grown by 6\%, in the simulation setting with the plant growth model.

For future work, the optimization problem can be extended into a model-predictive-control problem in a real world experiment. Our framework of collecting data, modelling the phenomena, and optimizing the various factors based on the developed model is generally flexible. We hope to extend this framework into optimizing the other aspects of the plant factory, which includes air-conditioning and nutrient delivery. It will also be interesting to consider other ancillary systems like solar PV or energy storage within the plant factory in the optimization problem.

\bibliography{lightning.bib}
\bibliographystyle{IEEEtran}

\end{document}